\documentclass[runningheads]{svmult}

\usepackage{graphicx}  % standard LaTeX graphics tool
\begin{document}
\title*{Post-Starburst Populations Near and Far -- The Potential of Near-IR
Spectroscopy}
\toctitle{Post-StarburstPopulations Near and Far -- The Potential
of Near-IR Spectroscopy} 
\titlerunning{Post-Starbursts in the Near-IR}
\author{Ariane Lan\c{c}on\inst{1}
\and Mustapha Mouhcine\inst{1}}
%\and Elsa Bertino\inst{1}}
%
\authorrunning{A. Lan\c{c}on \& M. Mouhcine}
\institute{Observatoire de Strasbourg (UMR\,7550), 11 rue de l'Universit\'e,
  F--67000 Strasbourg, France}
 
\maketitle              % typesets the title of the contribution
 
\begin{abstract}
The efficient use of near-IR data in studies of external stellar
populations depends on our ability to recognize the nature of the
predominant sources of light, and to interprete these findings in
terms of age and metallicity. Here we focus on elderly post-starburst
populations, with ages of $10^8 - 10^9$\,yrs. New models confirm
that they are indeed expected to display specific spectral signatures
in the near-IR,
due to variable M stars of the asymptotic giant branch and to 
carbon stars. The signatures depend on age and metallicity. 
We summarize the status of current quantitative predictions and
emphasize the importance of an empirical calibration of the spectral 
synthesis models.
\end{abstract}

\section{Overview} 

Near-IR data has been used very often to confirm the presence of an
\lq \lq evolved stellar component" in starburst galaxies. The need to
study stellar populations deeply embedded in dust is also frequently
invoked to justify near-IR observations. But what exactly are these cool
stellar populations seen between 1 and 3\,$\mu$m? Population
synthesis models tell us that red supergiants are the main sources
$10^7 - 10^8$\,yrs after a star formation episode; during the following
$10^9$\,yrs, stars of the upper, thermally pulsing asymptotic giant
branch (TP-AGB) become predominant; and finally giants of the early 
AGB (E-AGB) and of the first red giant branch (RGB) outweigh other
populations. The efficient use of near-IR data depends on our ability
to distinguish these categories from each other and to interprete
them in terms of age and metallicity. 

Very strong CO absorption (2.3\,$\mu$m) is a clear signature of a red 
supergiant population (\cite{BFP73} initiated the systematic
exploration of CO in cool stars; \cite{Retal80} analysed early
observations in starbursts; \cite{LRV96} discussed the 
diagnostic power of the feature and its limitations). But the dependence
of the red supergiant population on metallicity is strong
\cite{Massey98} and not reproduced by stellar evolution models unless
ad hoc corrections are invoked (\cite{LM95}, \cite{OGLSO99}). 
Therefore the threshold above which the CO absorption becomes
an unambiguous signature of a supergiant population
is uncertain. On the other hand, it has recently been claimed
that high quality, high spectral resolution near-IR observations allow 
for {\em relative} age-dating of young post-starburst populations to within 
$2-3$\,Myr, at least at solar metallicity (Gilbert, this workshop).

The focus of our work is on somewhat older post-starburst populations,
such as we are likely to find in evolved mergers or \lq \lq E+A"
galaxies, i.e. after the most obvious signs of a recent starburst have
faded away.
Between $10^8$ and $10^9$\,yrs of age, of the order of 50\,\% of the K band
light of a coeval population originates from TP-AGB stars.
Large stellar samples show that essentially all the TP-AGB stars are 
long period variables \cite{Wood99}. Spectral libraries as well as
model atmospheres demonstrate that this pulsation results in deeper
H$_2$O absorption bands than seen in E-AGB or RGB stars. This provides
us with a means of identifying strong contributions of oxygen-rich (e.g.
M type) TP-AGB stars. In addition, the convective dredge-up of carbon-rich
material from the stellar core into the envelope progressively turns some
TP-AGB stars into carbon stars. Those again have unmistakable spectral
signatures in the near-IR. Finally, TP-AGB stars may become dust
enshrouded mid-IR sources, and it has been suggested that the presence
of these modifies the location of stellar populations in near-IR
versus mid-IR two-colour diagrams \cite{BGS98}.
All the mentioned processes depend sensitively on the initial stellar
mass (and thus on post-starburst age) as well as on metallicity. This was the 
motivation for our work. We summarize its status below.

\section{Model construction and predictions}

Our population synthesis models are based on the code structure of
{\sc P\'egase} \cite{FRV97}, the stellar tracks of the \lq \lq Padova
Group" (\cite{BFBC93},\cite{FBBC94}) up to the end of the E-AGB,
and the spectral library of Lejeune et al. \cite{LCB97}.
We have completed this set of inputs with synthetic evolutionary
tracks for the TP-AGB that incorporate the effects of mass loss,
dredge-up and envelope burning of dredged-up materials (\cite{ML00},
\cite{ML01}); we also use a new library of cool stellar spectra \cite{LW00}.

\begin{figure}
\includegraphics[clip=,width=0.5\textwidth]{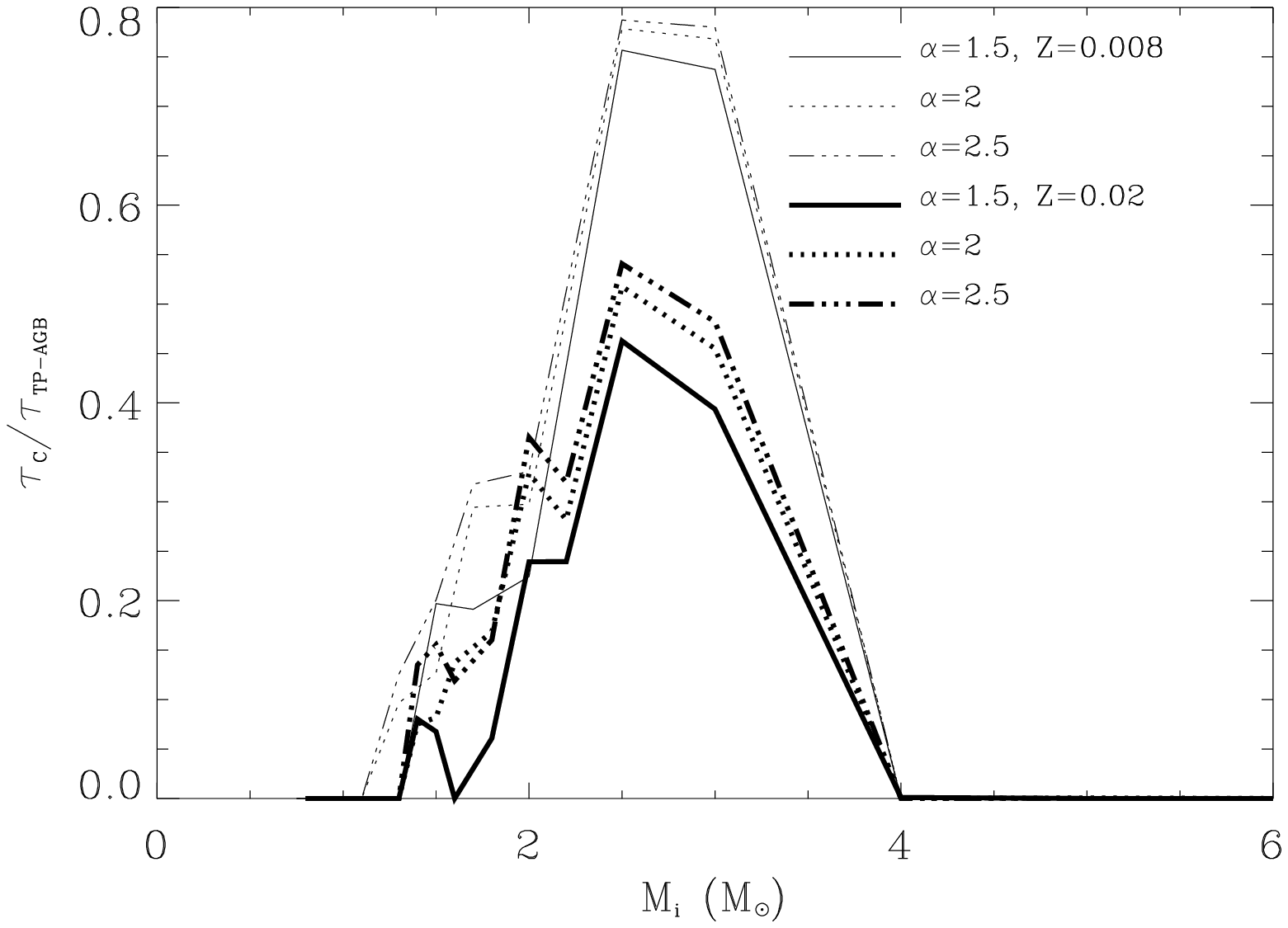}
\includegraphics[clip=,width=0.5\textwidth]{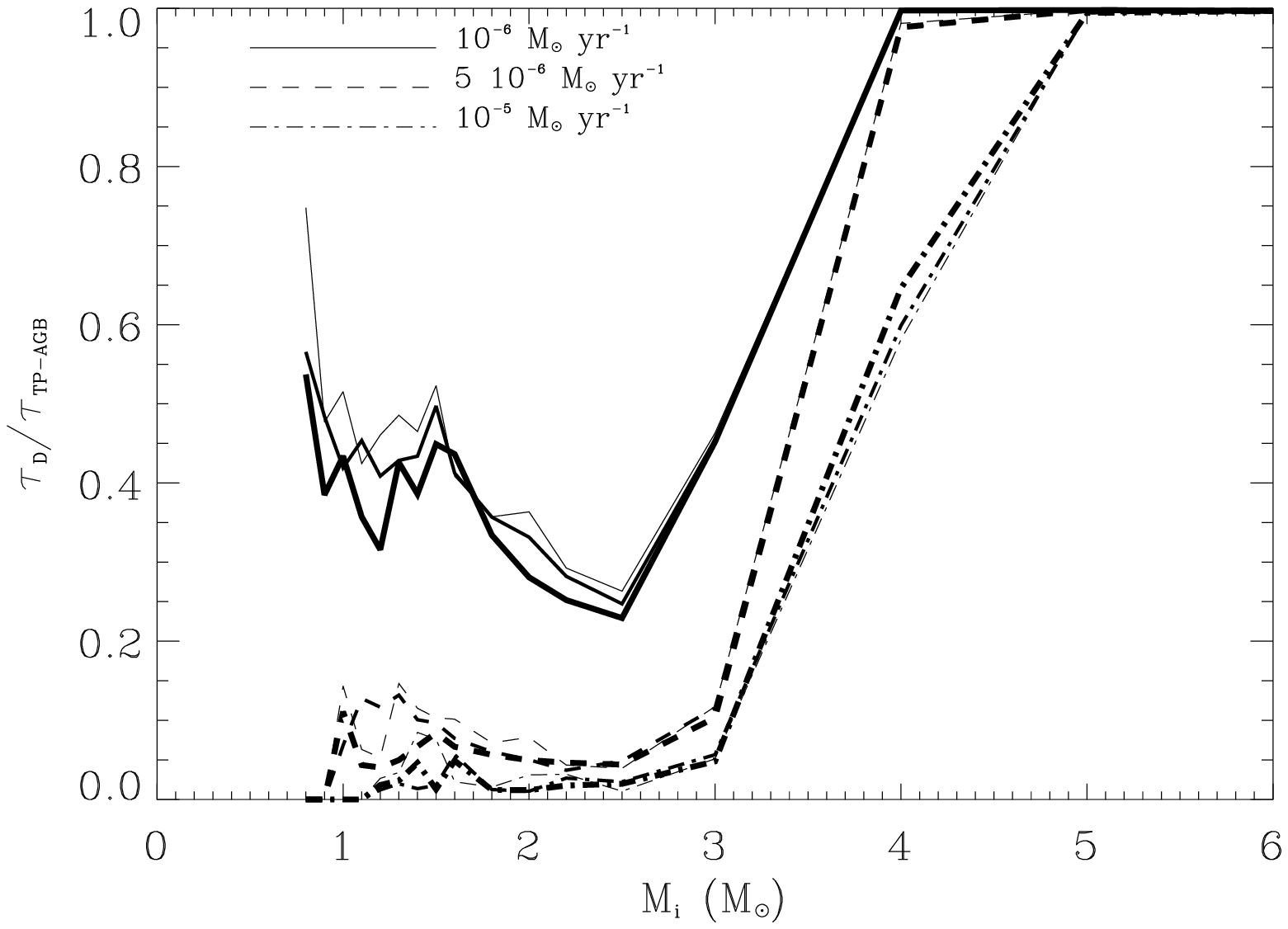}
\caption[]{
 {\em Left}\,: Fraction of the TP-AGB duration spent as a C-star, as 
 a function of initial stellar mass. Two metallicities and various
 values of the mixing length parameter are shown.
 {\em Right}\,: Fraction of the TP-AGB duration spent as a star with
 mass loss above $10^{-6}$, $5\times 10^{-6}$ and $10^{-5}$\,M$_{\odot}$/yr,
 at solar metallicity.
}
\label{fractions.fig}
\end{figure}

Various global properties can be determined without invoking
spectral libraries, on the basis of the TP-AGB tracks alone
\cite{ML01}. The values given here are based on our currently
prefered internal model parameters (e.g. mass loss prescription,
convective mixing length, dredge-up efficiency), which have been
tested against statistical properties in nearby stellar populations
(see below).
%; we note that further testing is still being done and refer to
%\cite{ML01} for details. 
The stars with the
longest TP-AGB durations (and thus the largest luminous energy output, 
integrated along the TP-AGB) are found to have initial masses ($M_i$) 
between 2 and 2.5\,M$_{\odot}$ (i.e. ages of 0.7 to 1.1\,Gyr);
stars with $M_i$ in the $1.7-3$\,M$_{\odot}$ range last 
$\sim 10^6$\,yrs or more in that phase. Mass loss has the strongest
direct influence on these durations, and itself depends on various
other model parameters through the instantaneous masses, luminosities
and radii. As the TP-AGB stars are the coolest
luminous stars present, the duration of that phase essentially controls the 
(V-K) colour of the stellar population \cite{GB98}.
The fraction of the TP-AGB duration spent as a carbon star is found to be 
maximal for objects with initial masses between 2.5 and 3\,M$_{\odot}$
(i.e. ages of 0.4 to 0.8\,Gyr), where it typically amounts 
$45\,\%$ at solar metallicity and about $80\,\%$ at the metallicity of the Large
Magellanic Cloud (Fig.\,\ref{fractions.fig}). 
Due to strong mass loss, stars above $M_i=3.5$\,M$_{\odot}$ have
relatively short TP-AGB durations ($< 5 \times 10^5$\,yrs) and are 
embedded in dust for a significant fraction of this time. This reduces
the predicted contribution of TP-AGB stars to the light at ages of
$0.1-0.4$\,Gyr, when compared to our earlier models \cite{LMFS99}.

As mentioned, our trust in the model prediction is based on a variety of tests.
For instance, the new evolutionary models for the TP-AGB successfully 
reproduce the empirical initial to final mass relation for stars in the 
solar neighbourhood. The predicted contributions of C-stars to the
near-IR light of intermediate age populations are consistent with
those observed in the Magellanic Clouds (e.g. \cite{PACFM83}).
One of us (M.M.) has coupled the TP-AGB models with 
chemical evolution models for galaxies (including the yields of 
intermediate mass stars as given by the new tracks). The tight relations 
observed in the Local Group galaxies between metallicity and the mean 
luminosity of C-stars, or between metallicity and the C-star/M-star number 
ratio can be reproduced very naturally. These results will be discussed
in detail elsewhere. 

Stellar libraries are required for the prediction of near-IR colours
or spectral features. Based on the extremely diverse AGB star spectra of 
\cite{LW00} 
($\lambda/\delta \lambda \simeq 1100$ between 1 and 2.5\,$\mu$m), 
we have constructed a regular sequence
of averages that can be more conveniently coupled with population
synthesis models \cite{LM01}. As a result, we predict that the signatures
of upper AGB stars are seen in the integrated light of a stellar populations
between $\sim 0.2$ and $\sim 1.2$\,Gyr after a starburst (\cite{ML01},
\cite{LMFS99}). The main signatures are extremely broad, and
best detected with low resolution data with very wide spectral coverage
(the K band spectrum alone is not sufficient). As the features reach accross
telluric absorption bands, observations at good IR sites are essential.
The features are due to H$_2$O if the TP-AGB stars are O-rich, to C$_2$ and
CN if they are C-rich. Narrow band indices can be defined that
detect TP-AGB stars independently of their chemical nature,
and others that then separate the two classes \cite{LMFS99}. Photometry
with an accuracy better than 5\,\% must be aimed for. We are 
exploring the feasibility of surveys with the available filters on
the NICMOS camera onboard the Hubble Space Telescope.

\section{The need for empirical calibration}

\begin{figure}
\includegraphics[clip=,width=0.43\textwidth]{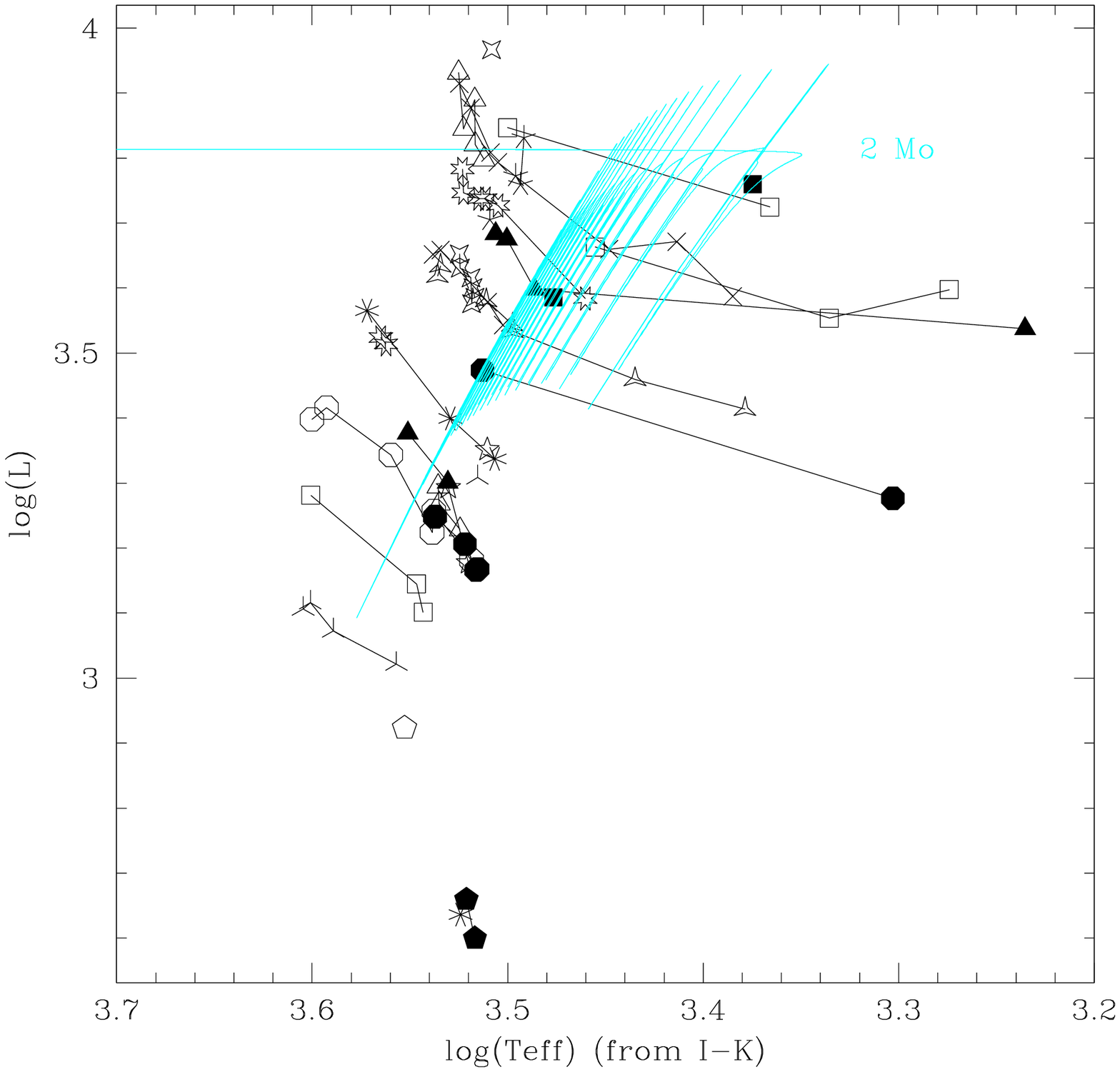}
\includegraphics[clip=,angle=90,width=0.56\textwidth]{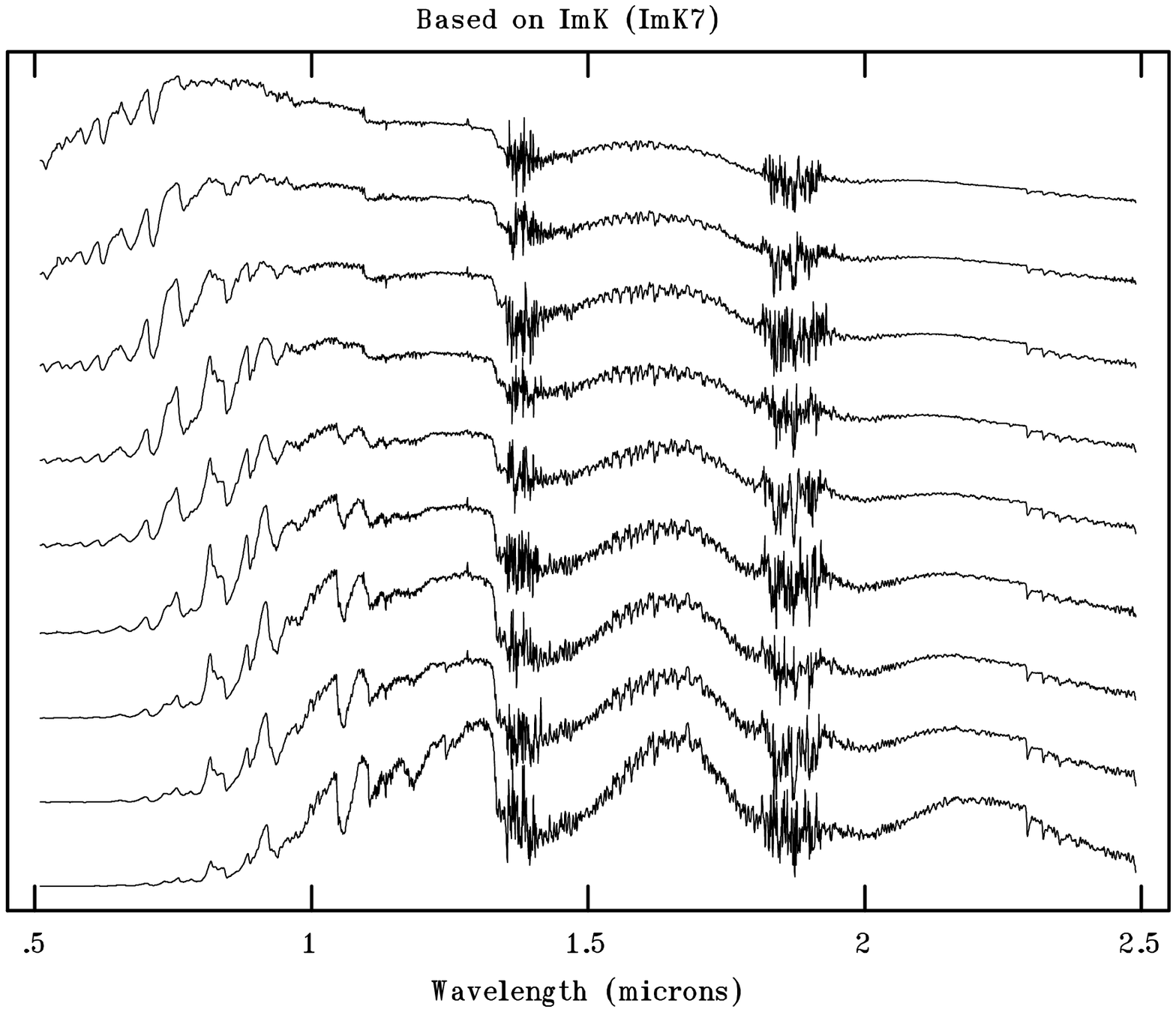}
\caption[]{{\em Left}\,: Estimated location of individual spectra of \cite{LW00}
in the HR diagram. Multiple observations of individual stars are
connected with lines. An illustrative TP-AGB track for a 2\,M$_{\odot}$ star
is also shown. {\em Right}\,: Empirical sequence of average O-rich TP-AGB 
star spectra \cite{LM01}.}
\label{HR.fig}
\end{figure}

Once TP-AGB evolutionary tracks have been constructed and tested,
only a part of the work is 
done: associating a spectrum with a given point of the theoretical
HR diagram is difficult, as illustrated in Fig.\,\ref{HR.fig}. 
Individual stars vary significantly in luminosity and colour
temperature. While optical colours and temperature indicators
such as (V-K) correlate well, the near-IR properties (colours,
strength of the H$_2$O or CO bands) are very dispersed at a
given colour temperature. In addition, no model atmospheres are
available to reliably relate an instantaneous colour temperature
to the effective temperature of the simple static stars on which
evolutionary tracks are based. The regularity of the spectral sequence
obtained by \cite{LM01} gives some confidence in the evolution of
the spectra with temperature; but only empirical calibrations, based
on stellar populations with ages and metallicities known from
optical observations, can constrain absolute relationships. 

\begin{figure}
\includegraphics[clip=,angle=90,width=0.5\textwidth]{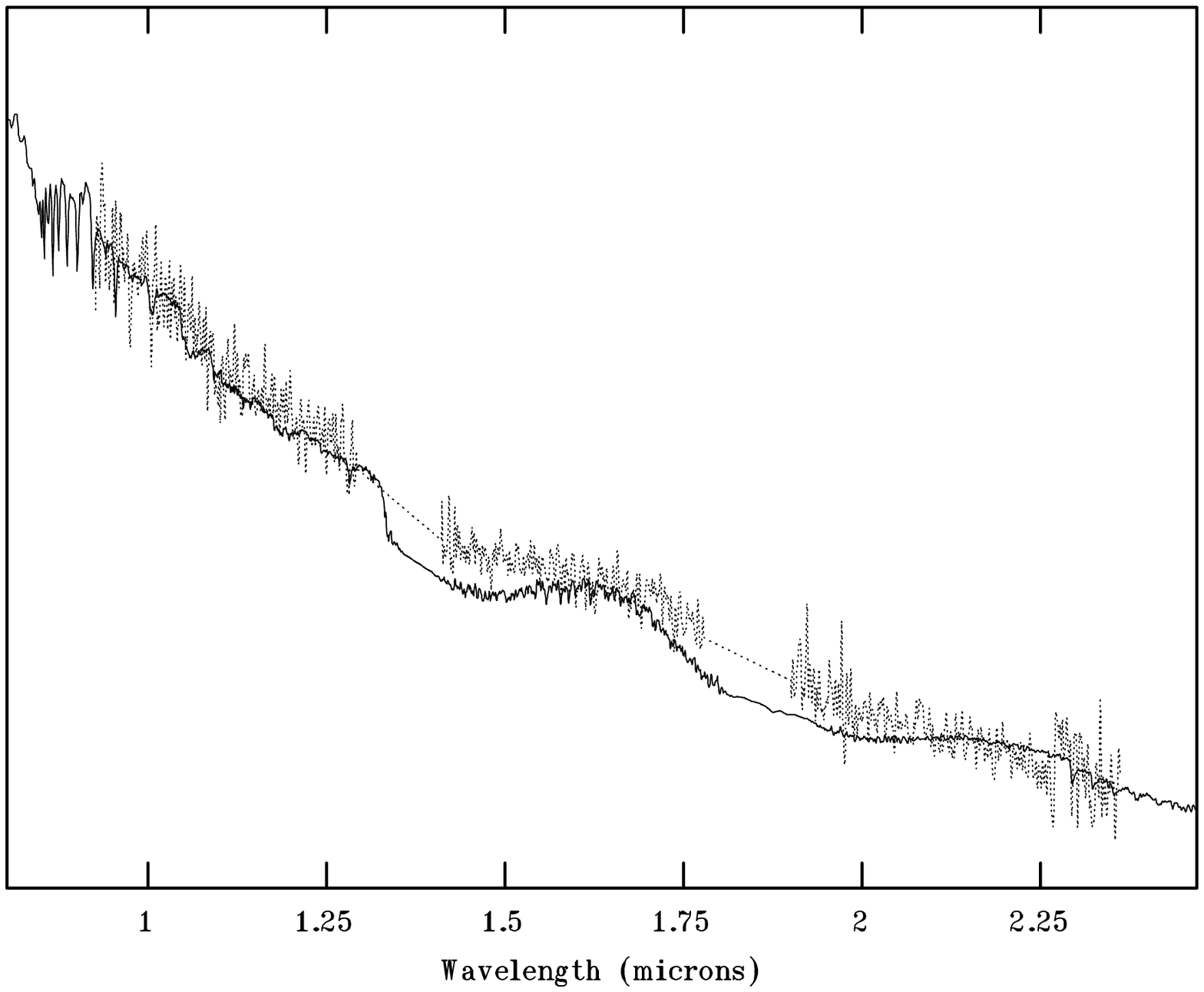}
\includegraphics[clip=,angle=90,width=0.5\textwidth]{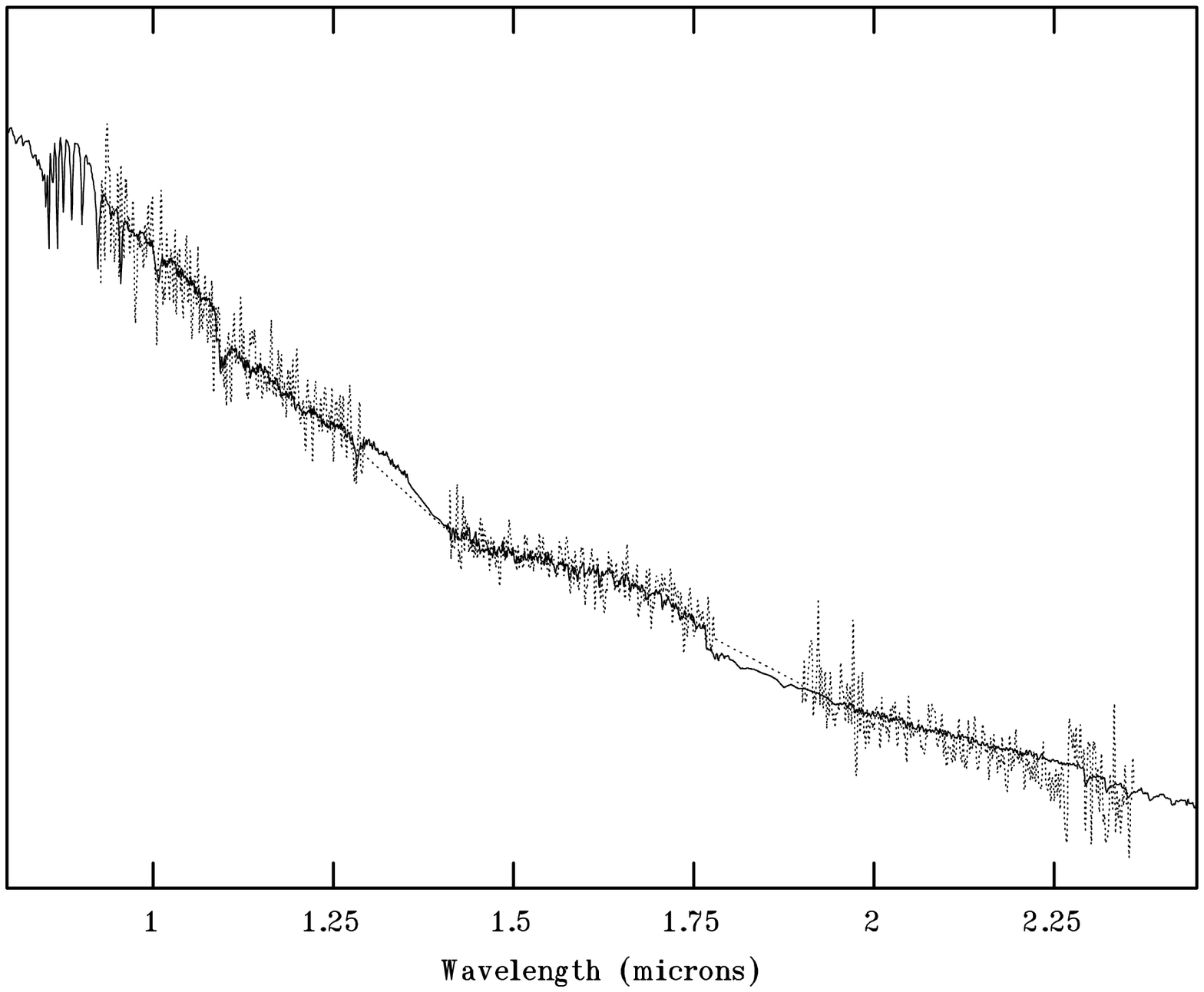}
\caption[]{The observed spectrum of W3 and two illustrative models.
{\em Left}\,: solar metallicity, formation of C-stars switched off,
cool TP-AGB temperature scale. This model is ruled out. {\em Right}\,:
LMC metallicity, formation of C-stars allowed. This model matches the
data well.}
\label{fits.fig}
\end{figure}

Despite the more numerous near-IR instruments now available on large
telescopes, appropriate targets for the model calibration
remain hard to find. The LMC and M33
clusters, of which several have adequate ages, tend to be too small:
they contain a handful of TP-AGB stars and their integrated properties
are affected by the stochastic nature of this subpopulation. In \cite{LW00},
it was shown that stellar populations with total masses of 
$\sim 10^5$\,M$_{\odot}$ are needed to obtain significant
constraints on the TP-AGB models.
% (fig.\,\ref{fluct.fig}). 
Massive stellar clusters around
merging galaxies are well suited, but only a few are bright enough
for near-IR spectroscopy. We have observed the 500\,Myr old cluster W3
in NGC\,7252 with SOFI (NTT, ESO), between 1 and 2.4\,$\mu$m. Illustrative
adjustments with models are shown in Fig.\,\ref{fits.fig}. The figures 
illustrate the need for broad wavelength coverage and excellent
relative calibrations of the various atmospheric windows (with SOFI
the overlap between grisms ensures this). Our preliminary conclusions
are the following. The data reject models that have few C-stars
{\em and} use a cool effective temperature scale for the TP-AGB
spectra: the water bands in those are too strong. This suggests that
the metallicity is not more than solar and that the coolest of the
Mira temperature scales in the literature is inappropriate for our
purpose. Good fits to the spectral shape
are obtained at solar and LMC metallicities, with extinction values
compatible with optical estimates \cite{SS98}.  
More calibration points will be required, but this first test
already restricts the allowed range of parameters and gives us
confidence in the new tool.

\section{Conclusion}

Based on new synthetic evolution models for the TP-AGB and on a 
purpose designed library of stellar spectra, we have shown that
near-IR spectra of post-starburst populations indeed carry 
information about post-starburst age and metallicity. Elderly
post-starbursts with ages of $10^8$ to $10^9$\,yrs can 
be distinguished from younger or older ones, using the specific
spectral signatures of O-rich and C-rich TP-AGB stars. The
chemical nature of the predominant TP-AGB stars is a metallicity
indicator. The population of dust-obscured TP-AGB stars depends
both on age and metallicity. We have pointed out
that the relevant spectral features can be detected at low spectral 
resolution as long as 
a broad continuous spectral coverage is obtained. High quality narrow 
band photometry can also be used. The quantitative
prediction of the strength of the near-IR features must be
based on empirical calibrations. We have described the current 
status of these.

Applications of this work include studies of the duration of 
star formation episodes in starburst galaxies or mergers, of
the survival timescale of young starburst clusters in the
galaxy environment, and of the nature of ``E+A" galaxies.
First observations of ``E+A" galaxies will be available to us soon.

%INDEX%%%%%%%%%%%%%%%%%%%%%%%%%%%%%%%%%%%%%%%%%%%%%%%%%%%%%%%%%%%%%%%
% Please check with the editor of your book whether he plans to
% include a "mutual" subject index - if so, please code your entries
% in the standard syntax. For your own purposes you may print your
% "personal" index by using the following commands:
%
%\clearpage
%\addcontentsline{toc}{section}{Index}
%\flushbottom
%\printindex
%%%%%%%%%%%%%%%%%%%%%%%%%%%%%%%%%%%%%%%%%%%%%%%%%%%%%%%%%%%%%%%%%%%%%

\end{document}